\documentclass{article}
\usepackage{amsmath}
\usepackage{graphicx}
\usepackage{color}
\usepackage{ulem}
\usepackage{amsfonts}
\usepackage{epsfig}

\newcommand{\R}{\mbox{$I \kern -4pt R$}}

\begin{document}

\title{Spiral Instabilities in Rayleigh-B\'enard Convection under Localized
Heating}
\author{M. C. Navarro$^1$, A. M. Mancho$^2$and H. Herrero$^1$ \\
{$^1$ {\small Departamento de Matem\'aticas}} {\small Facultad de Ciencias Qu%
\'{\i}micas}\\
{\small Universidad de Castilla-La Mancha} {\small 13071 Ciudad Real, Spain.}%
\\
{$^2$ {\small Instituto de Matem\'aticas y F\'{\i}sica Fundamental}}\\
{\small Consejo Superior de Investigaciones Cient\'{\i}ficas}\\
{\small Serrano 121, 28006 Madrid, Spain.}}
\date{}
\maketitle

\begin{abstract}
We study from the numerical point of view, instabilities developed in a
fluid layer with a free surface, in a cylindrical container which is
non-homogeneously heated from below. In particular we consider the case in
which the applied heat is localized around the origin. An axysimmetric basic
state appears as soon a non-zero horizontal temperature gradient is imposed.
The basic state may bifurcate to different solutions depending on vertical
and lateral temperature gradients and on the shape of the heating function. We find
different kinds of instabilities: extended patterns growing on the whole
domain which include those known as targets and spirals waves. Spirals are
present even for infinite Prandtl number. Localized structures both at the
origin and at the outer part of the cylinder may appear either as Hopf or
stationary bifurcations. An overview of the developed instabilities as a
function of the dimesionless parameters is reported in this article.
\end{abstract}

\noindent Corresponding author: Ana M. Mancho, Instituto de Matem\'aticas y F%
\'{\i}sica Fundamental, Consejo Superior de Investigaciones Cientificas,
Serrano 121, 28006, Madrid. Phone: +34 91 5616800 ext 2408. Fax: +34 91
5854894. e-mail: A.M.Mancho@mat.csic.es

{\bf Spiral and target waves instabilities in Rayleigh-B\'ernard convection were experimentally found
in early nineties and several theoretical mechanisms have been proposed to justify them.
In this article we prove that these and other structures may be understood as the linearly growing modes
of non trivial basic states generated by a non uniform heating function. In this framework we find 
that also localized structures may appear. These patterns are localized at different positions in the domain depending on 
the basic flow from which they bifurcate. This article  reports a sytematical numerical 
study  on these transitions.}

\section{Introduction}

Instabilities and pattern formation in buoyant flows have been extensively
studied in recent years. Classically, heat is applied uniformly from below 
\cite{bern} and the conductive solution {(also called the basic solution)}
becomes unstable for {a critical vertical} temperature gradient beyond a
certain threshold. Typically, instabilities in Rayleigh-B\'{e}nard
convection lead to roll patterns, however spiral waves have been also
reported, in both theoretical and experimental results. First
experimental observations on spirals \cite{boden1,morris1} in Rayleigh-B\'{e}%
nard convection were rather surprising as they were carried out for a set of
external parameter values where parallel rolls should be stable. In
experiments, spirals have been reported for finite Prandtl number \cite%
{boden2} and both single and multiarmed giant spirals and spiral chaos have
been observed. In \cite{nature} transitions between spiral and target states
are reported. First theoretical studies attempting to justify spiral
patterns in Rayleigh B\'{e}nard convection were performed just in model
equations such as Swift-Hohenberg type equations \cite{best,gunton}. These
results were confirmed by later simulations of the full governing equations 
\cite{pesch,pesch2}.

A more general setup for Rayleigh-B\'{e}nard convection than the case of
uniform heating, consists of a basic dynamic flow imposed by a non-zero
horizontal temperature gradient which may be either constant or not.
Numerical results obtained at the former setup in an annular domain address
the importance of both vertical and horizontal temperature
gradients in the development of instabilities \cite{pof,jpa,brief,pof2}. In
this work we consider a cylidrical domain with a localised Gaussian-like
heating at the bottom. Therefore, apart from vertical and horizontal
temperature gradients, a heat parameter related to the  shape of the
heating profile at the bottom is introduced. Our results show that many
different types of instabilities may appear depending on the basic flow and
external parameters. We study how instabilities depend on 
dimensionless parameters and we recover many results as those described for
the classical uniform heating case. Giant single armed spiral waves appear
and their instability threshold is close to those patterns refered as
targets. Localised and extended spirals are identified and both of them are
found for finite and infinite Prandtl number. Moreover, stationary patterns
appear both with local and extended structure.

The paper is organized as follows. Section 2 describes the physical
setup, the general mathematical formulation of the problem in a
dimensionless form, for basic solutions and their linear stability analysis.
Section 3 explains the numerical method in detail and presents a test of
convergence. In section 4 the numerical results on both, the basic and
growing perturbations, are discussed and conclusions are presented.

\section{ Formulation of the problem}

The physical set up (see figure 1) consists of a horizontal fluid layer in a
cylindrical container of radius $l$ ($r$ coordinate) and depth $d$ ($z$
coordinate). The upper surface is open to the atmosphere and the bottom
plate is rigid. At $z=0$ the imposed temperature is a Gaussian profile which
takes the value $T_{\max }$ \ at $r=0$ \ and the value $T_{\min }$ \ at the
outer part ($r=l$). The environmental temperature is $T_{0}$. We define $%
\triangle T_{v}=T_{\max }-T_{0}$, $\triangle T_{h}=T_{\max }-T_{\min }$ and $%
\delta =\triangle T_{h}/\triangle T_{v}$.

In the equations governing the system $u_{r},$ $u_{\phi }$ and $u_{z}$ are
the components of the velocity field $u$, $T$ is the temperature, $p$ is the
pressure, $r$ is the radial coordinate and $t$ is the time. The magnitudes
are expressed in dimensionless form after rescaling in the following form: $
{\bf r}^{\prime }={\bf r}/d,$ $t^{\prime }=\kappa t/d^{2},$ $u^{\prime }=du/\kappa ,$ $%
p^{\prime }=d^{2}p/\left( \rho _{0}\kappa \nu \right) ,$ $\Theta =\left(
T-T_{0}\right) /\triangle T$. Here ${\bf r}$ is the position vector, $\kappa $ the thermal diffusivity, $%
\nu $ the kinematic viscosity of the liquid and $\rho _{0}$ is the mean
density at the environment temperature $T_{0}$. After reescaling the domain $%
\Omega _{1}=[0,l]\times \lbrack 0,d]$ is transformed into $\Omega
_{2}=[0,\Gamma ]\times \lbrack 0,1]$ where $\Gamma =l/d$ is the aspect ratio

The system evolves according to the momentum and the mass balance equations
and to the energy conservation principle, which in dimensionless form are
(the primes in the corresponding fields have been dropped), 
\begin{eqnarray}
&& \nabla \cdot u =0,  \label{1general} \\
&& \partial _{t}\Theta +u\cdot \nabla \Theta =\nabla^2 \Theta,
\label{3general} \\
&& \partial _{t}u+\left( u\cdot \nabla \right) u =\mathrm{Pr}\left( - \nabla
p+\nabla^2 u +\mathrm{R} \Theta e_{z}\right),\,\,\,\,\,  \label{2general}
\end{eqnarray}
where the operators and fields are expressed in cylindrical coordinates and
the Oberbeck-Bousinesq approximation has been used. Here $e_{z}$ is the unit
vector in the $z$ direction. The following dimensionless numbers have been
introduced: the Prandtl number Pr$=\nu/\kappa $ and the Rayleigh number R$%
=g\alpha \triangle Td^{3}/\kappa \nu $, which represents the buoyant effect.
In these definitions $\alpha$ is the thermal expansion coefficient and $g$
is the gravity constant.

We discuss now the boundary conditions (bc). The top surface is flat, which
implies the following condition on the velocity, 
\begin{equation}
u_{z}=0,\;\;\mbox{on}\;\;z=1.  \label{eqbc1}
\end{equation}%
and free slip, 
\begin{equation}
\partial _{z}u_{r}=0,\;\;\partial _{z}u_{\phi }=0,\;\;\mbox{on}\;\;z=1.
\label{eqbc2}
\end{equation}%
%
%
Lateral and bottom walls are rigid, so 
\begin{eqnarray}
u_{r} &=&u_{\phi }=u_{z}=0,\;\;\mbox{on}\;\;z=0,  \label{eq:bcz0} \\
u_{r} &=&u_{\phi }=u_{z}=0,\;\;\mbox{on}\;\;r=\Gamma .  \label{eq:bcr11}
\end{eqnarray}%
For temperature we consider the dimensionless form of Newton's law for heat
exchange at the surface, 
\begin{equation}
\partial _{z}\Theta =-B\Theta ,\;\;\mbox{on}\;\;z=1,  \label{biot}
\end{equation}%
where $B$ is the Biot number. At the bottom a gaussian profile is imposed, 
\begin{equation}
\Theta =1-\delta (e^{{\large (}\frac{1}{\beta }{\large )}^{2}}-e^{(\frac{1}{%
\beta }-(\frac{x}{\Gamma })^{2}\frac{1}{\beta })^{2}})/(e^{(\frac{1}{\beta }%
)^{2}}-1)\;\;\mbox{on}\;\;z=0,  \label{10}
\end{equation}%
{where $\beta $ is a measure of the sharpness of the profile. In
figure 2 several plots of this profile for different values of the
parameters can be seen}. The lateral wall is insulating, 
\begin{equation}
\partial _{r}\Theta =0\;\;\;\mbox{on}\;\;\;r=\Gamma .  \label{11}
\end{equation}%
The use of cylindrical coordinates, which are singular at $r=0$, 
requires regularity conditions on velocity, pressure and temperature
fields. In general, these conditions are expressed as follows \cite{numerico}%
\begin{equation}
\frac{\partial (u_{r}e_{r}+u_{\phi }e_{\phi })}{\partial {\phi }}=\partial
_{\phi }u_{z}=\partial _{\phi }p=\partial _{\phi }\Theta =0\;\;\mbox{on}%
\;\;r=0,  \label{regularity}
\end{equation}%
{where $e_{r}$ and $e_{\phi }$ are the unit vectors in the $r$ and $\phi $
directions respectively}. To summarize, the dimensionless equations contains
these external parameters $(\mathrm{R},\Gamma ,\mathrm{Pr},\delta ,B,\beta )$. 
Our aim is to describe the type of bifurcations that appear
when some of those dimensionless quantities are varied.

\subsection{ Basic state and linear stability analysis}

The horizontal temperature gradient at the bottom (i.e, $\delta \neq 0$)
settles a stationary convective motion which is called basic state. It is a
time independent solution to the stationary problem obtained from  equations
(\ref{1general}-\ref{2general}). The basic state is axisymmetric
therefore it depends only on $r-z$ coordinates (i.e. all $\phi $ derivatives
are zero). The velocity field of the basic flow is restricted to $%
u=(u_{r},u_{\phi }=0,u_{z})$. Regularity conditions (\ref{regularity}) now
become 
\begin{equation}
u_{r}=\partial _{r}u_{z}=\partial _{r}p=\partial _{r}\Theta =0\;\;\mbox{on}%
\;\;r=0.  \label{regularityb}
\end{equation}%
We have solved numerically the simplified equations for the basic state
together with its boundary conditions. We use a Chebyshev collocation method
with details given in section 3.


The stability of the basic state is studied by perturbating it with a vector
field depending on the $r,\phi $ and $z$ coordinates, in a fully 3D analysis:%
\begin{eqnarray}
u_{r}(r,\phi ,z) &=&u_{r}^{b}(r,z)+\bar{u}_{r}(r,z)\exp (ik\phi +\lambda t),
\label{(3.1)} \\
u_{\phi }(r,\phi ,z) &=&u_{\phi }^{b}(r,z)+\bar{u}_{\phi }(r,z)\exp (ik\phi
+\lambda t),  \label{(3.2)} \\
u_{z}(r,\phi ,z) &=&u_{z}^{b}(r,z)+\bar{u}_{z}(r,z)\exp (ik\phi +\lambda t),
\label{(3.3)} \\
\Theta (r,\phi ,z) &=&\Theta ^{b}(r,z)+\overline{\Theta }(r,z)\exp (ik\phi
+\lambda t),  \label{(3.4)} \\
p(r,\phi ,z) &=&p^{b}(r,z)+\overline{p}(r,z)\exp (ik\phi +\lambda t).
\label{(3.5)}
\end{eqnarray}%
Here the superscript $b$ indicates the corresponding quantity in the basic
state and the bar refers to the perturbation. We have considered Fourier
mode expansions in the angular direction, because along it  boundary
conditions are periodic. Expressions (\ref{(3.1)}-\ref{(3.5)}) are replaced
into basic equations (\ref{1general}-\ref{2general}) and the resulting
system is linearized. Boundary conditions for  perturbations $(\bar{u}%
_{r},\bar{u}_{\phi },\bar{u}_{z},\overline{\Theta },\overline{p})$ are found
by substituting (\ref{(3.1)}-\ref{(3.5)}) into (\ref{eqbc1}-\ref{regularity}%
). Regularity conditions (\ref{regularity}) depend now on the wavenumber $k$%
: 
\begin{eqnarray}
&&\bar{u}_{r}=\bar{u}_{\phi }=\frac{\partial \bar{u}_{z}}{\partial r}=\frac{%
\partial \bar{\Theta}}{\partial r}=\frac{\partial \bar{p}}{\partial r}=0,\;\;%
\mbox{for}\;\;k=0,  \label{eq:bcr0m0} \\
&&\bar{u}_{r}+i\bar{u}_{\phi }=\bar{u}_{z}=\bar{\Theta}=\bar{p}=0,\;%
\mbox{for}\;\;k=1,  \label{eq:bcr0m1} \\
&&\bar{u}_{r}=\bar{u}_{\phi }=\bar{u}_{z}=\bar{\Theta}=\bar{p}=0,\;\mbox{for}%
\;\;k\neq 0,\,1.  \label{eq:bcr0}
\end{eqnarray}%
The resulting problem is an eigenvalue problem in $\lambda $. If $Re(\lambda
)<0$ for all eigenvalues the basic state is stable while if 
there exists a value of $\lambda $  such as $Re(\lambda )>0$ the
basic state becomes unstable. The condition $Re(\lambda )=0$ may be
satisfied for certain values of the external parameters, $(\mathrm{R},\Gamma
,\mathrm{Pr},\delta ,B,\beta )$, which define the critical threshold. At the
critical threshold, a stationary bifurcation takes place if $Im(\lambda )=0$
while it is a Hopf bifurcation if $Im(\lambda )\neq 0$.

\section{Numerical method}

We have solved both the basic state and the linear stability problem as
stated in (\ref{1general}-\ref{2general}), i.e. in primitive variables
formulation by expanding the fields with Chebyshev polynomials (see Ref. 
\cite{bm}). Using this technique, the problem of the spurious modes for
pressure arises \cite{bm,canuto}, which we have solved using the method
proposed in \cite{numerico}, taking additional boundary conditions. They are
obtained by the continuity equation at $z=0$ and the normal component of the
momentum equations on $r=\Gamma $ and $z=1$, 
\begin{eqnarray}
&&\nabla \cdot u=0,\;\;\mbox{on}\;\;z=0,  \label{bcespecial1} \\
&&\mathrm{Pr}^{-1}\left( \frac{\partial u_{r}}{\partial t}+u_{r}\frac{%
\partial u_{r}}{\partial r}+\frac{u_{\phi }}{r}\frac{\partial u_{r}}{%
\partial \phi }+u_{z}\frac{\partial u_{r}}{\partial z}-\frac{u_{\phi }^{2}}{r%
}\right) \overset{}{=}  \notag \\
&&\overset{}{=}-\frac{\partial p}{\partial r}+\Delta u_{r}-\frac{u_{r}}{r^{2}%
}-\frac{2}{r^{2}}\frac{\partial u_{\phi }}{\partial \phi },\;\;\mbox{on}%
\;\;r=\Gamma ,  \label{bcespecial2} \\
&&\mathrm{Pr}^{-1}\left( \frac{\partial u_{z}}{\partial t}+u_{r}\frac{%
\partial u_{z}}{\partial r}+\frac{u_{\phi }}{r}\frac{\partial u_{z}}{%
\partial \phi }+u_{z}\frac{\partial u_{z}}{\partial z}\right) \overset{}{=} 
\notag \\
&&\overset{}{=}-\frac{\partial p}{\partial z}+\Delta u_{z}+\mathrm{R}\Theta
,\;\;\mbox{on}\;\;z=1,  \label{bcespecial3}
\end{eqnarray}%
where $\Delta =r^{-1}\partial /\partial r(r\partial /\partial
r)+r^{-2}\partial ^{2}/\partial \phi ^{2}+\partial ^{2}/\partial z^{2}$.

\subsection{Basic state}

We have solved numerically the stationary axisymmetric version of equations (%
\ref{1general}-\ref{2general}) together with the boundary conditions, by
treating the nonlinearity with a Newton-like iterative method. In a first
step the nonlinearity was neglected and a solution was found by solving the
linear system: $u_{r}^{0}$, $u_{\phi }^{0},$ $u_{z}^{0}$, $p^{0}$, $\Theta
^{0}$. This solution was corrected by perturbation fields: $%
u_{r}^{1}=u_{r}^{0}+\bar{u}_{r}$, $u_{z}^{1}=u_{z}^{0}+\bar{u}_{z}$, $%
p^{1}=p^{0}+\bar{p}$ and $\Theta ^{1}=\Theta ^{0}+\bar{\Theta}$. These
expressions are introduced into the equations which are linearized around
the approach at step $0$. This procedure is repeated with the new
solution and so on. The criterion of convergence considered to stop the
iterative procedure is that the $l^{2}$ norm of the computed perturbation
should be less than $10^{-9}.$

At each step the resulting linear system for perturbations is solved by
expanding any unknown perturbation field \textbf{x} in Chebyshev
polynomials, 
\begin{equation}
\displaystyle\mathbf{x}=\sum_{l=0}^{L-1}\sum_{n=0}^{N-1}a_{ln}^{\mathbf{x}}{T%
}_{l}(r){T}_{n}(z).  \label{chebexp}
\end{equation}%
For computational convenience the domain $\Omega _{2}=\left[ 0,\Gamma \right]
\times \left[ 0,1\right] $ is transformed into $\Omega =\left[ -1,1\right]
\times \left[ -1,1\right] $. This change of coordinates introduces scaling
factors in equations and boundary conditions which are not explicitly given
here. There are $4\times L\times N$ unknows which are determined by a
collocation method. In particular, expansions (\ref{chebexp}) are replaced
into the linearized, stationary, axisymmetric equations and
boundary conditions, and those are posed at the Gauss-Lobatto collocation
points $(r_{j},z_{i})$, 
\begin{eqnarray}
r_{j} &=&\cos \left( \left( \frac{j-1}{L-1}-1\right) \pi \right)
,\,\,\,\forall \,\,j=0...L. \\
z_{i} &=&\cos \left( \left( \frac{i-1}{N-1}-1\right) \pi \right)
,\,\,\,\forall \,\,i=0...N.
\end{eqnarray}%
Evaluation rules are as follows: the conveniently simplified Eqs. (\ref%
{1general}-\ref{2general}) are evaluated at nodes $%
i=2,...,N-1,\;j=2,...,L-1$; the boundary conditions at $z=-1$, (\ref{eq:bcz0}%
) and (\ref{10}) at $i=1,\;j=2,...,L-1$; the boundary conditions at $z=1$, (%
\ref{eqbc1}-\ref{eqbc2}) and (\ref{biot}) at $i=N,\;j=2,...,L-1$ ; the
boundaries at $r=-1$, (\ref{regularityb}) at $i=1,...,N,\;j=1$; finally the
boundaries at $r=1$, (\ref{eq:bcr11}) and (\ref{11}) at $i=1,...,N,\;j=L$.
The system of equations is completed with additional boundary conditions
that eliminate spurious modes for pressure. At $z=-1$ equation (\ref%
{bcespecial1}) is evaluated at nodes $i=1,\;j=2,...,L-1$; at $z=1$ equation (%
\ref{bcespecial3}) is evaluated at nodes $i=N,\;j=2,...,L$; at $r=1$
equation (\ref{bcespecial2}) is evaluated at $i=1,...,N-1,\;j=L$. With these
rules the matrix associated to the linear algebraic system is singular, due
to the fact that pressure is defined up to an additive constant. To fix this
constant the boundary condition (\ref{bcespecial2}) in node $i=N-2,j=L$ is
replaced by a Dirichlet condition for pressure ($i.e$, $p=0$ at $i=N-2,j=L$%
). In this way at each iteration step a linear system of the form $AX=B$
is obtained, where $X$ is a vector containing $4\times
L\times N$ unknowns and $A$ is full rank matrix of order $P\times P$ with $%
P=4\times L\times N$. This can be easily solved with standard routines.

\subsection{Linear stability analysis}

The eigenvalue problem described in section 3.1 is discretized by
expanding pertubations (\ref{(3.1)}-\ref{(3.5)}) in a truncated series of
orthonormal Chebyshev polynomials as we did with the basic state. These
expressions 
are replaced into the linearized version of equations (\ref{1general}-\ref%
{2general}) and boundary and regularity conditions (\ref{eqbc1}-\ref%
{regularity}). We use as in the previous subsection a collocation method
where equations are evaluated at the Gauss-Lobatto points. Evaluation rules
are analogous except that now regularity conditions are replaced by those
expressed in (\ref{eq:bcr0m0}-\ref{eq:bcr0}). We notice that for $k=1$ there
are only four boundary conditions at $r=-1$, and therefore in order to fit
the unknowns with the equations we have diminished the expansion range of
the $u_{\phi }$ field by one order, 
\begin{equation*}
\displaystyle u_{\phi }=\sum_{l=0}^{L-2}\sum_{n=0}^{N-1}a_{ln}^{v}\mathrm{T}%
_{l}(r)\mathrm{T}_{n}(z).
\end{equation*}%
In this way for $k=1$ we have $P=4\times N\times L+N\times (L-1)$ unknowns
and equations while these are $P=5\times L\times N$ for $k>1$ and $P=4\times
L\times N$ for $k=0$. For the eigenvalue problem these rules are explicitly
detailed in Ref. \cite{numerico} with the sole difference that here
equations are linearized around a non trivial basic state instead of around
a conductive solution.

The eigenvalue problem is then transformed into its discrete form 
\begin{equation}
Aw=\lambda Bw  \label{(3.6)}
\end{equation}%
where $w$ is a vector which contains $P$ unknowns and $A$ and $B$ are $%
P\times P$ matrices. The discrete eigenvalue problem (\ref{(3.6)}) 
has a finite number of eigenvalues $\lambda _{i}$. The stability
condition explained in previous subsection must be required now upon $%
\lambda _{\max }$ where $\lambda _{\max }=\mathrm{max}Re(\lambda _{i})$. In
Ref. \cite{oxford} many details on how to solve this problem
efficiently are given.

As in \cite{jpa}, we have carried out a test on the convergence of the
method that let us assure the correctness of the results. Table I shows some
results on convergence rates. We find that expansions of order $33\times 9$
\ are enough to ensure accuracy within 1\%.

\section{ Numerical Results}

In this section we describe numerical results found for different values of
external parameters. The problem formulated through Eqs. (\ref%
{1general}-\ref{regularity}) depends on the dimensionless numbers $(\mathrm{R%
},\Gamma ,\mathrm{Pr},\delta ,B,\beta )$. In particular we are interested in
analyzing how those parameters related to heat conditions impose basic flows
leading to a great variety of instabilities. Some of them such as giant
spirals or targets have been already described in experiments for
the case of Rayleigh-B\'{e}nard convection with uniform heating \cite%
{boden1,boden2,nature}.

Issues related to horizontal and vertical temperature differences have
already been addressed in \cite{jpa,brief,pof2} but here we study for the
first time the influence of the shape parameter $\beta $. We restringe
ourselves to large and medium $\beta $ values, i.e. $\beta >0.5$.
In this region the numerical method has good convergence and among other
instabilities a giant spiral is observed. Smaller $\beta $ values will be
adressed in future works. We also analyze the influence of $\delta $
meassuring the quotient between lateral and vertical temperature gradients. 
Figure 2a) illustrates the dependence of the temperature boundary
condition on $\beta $ while figure 2b) shows its dependence on $\delta $.

\subsection{Basic states}

We have solved numerically the stationary axisymmetric version of equations (%
\ref{1general}-\ref{2general}) together with the boundary conditions, as
explained in section 3.1. Different basic state solutions are
obtained depending on the parameters. Figure 3a) displays a return
flow with clearly inverted temperature gradients and a single roll on the
velocity field. Figure 3b) shows a linear temperature field with no
inversions on the temperature gradient and several corotative rolls. 
Figures 3c) and 3d) displays intermediate states. In
Figure 3c) we see a rather vertical temperature gradients with
slight inversions and a slight corrotative roll. Figure 3d)
displays a return flow where the outer part of the cylinder is hotter than 
the inner one. These basic flows become unstable to
different spatial structures as we explain next.

\subsection{Instabilities}

We have studied numerically the linear stability of the numerical basic 
states following explanations of section 3.2. We consider
the external parameter R as control parameter. By control parameter we mean
the parameter that changed leads to an instability while all the others are
fixed. This occurs when $\lambda _{\max }(\mathrm{R})$ changes from a
negative value to a positive one as R varies. The value of R$_{c}$ for which 
$\lambda _{\max }(\mathrm{R}_{c})=0$ \ is the critical value. Figure 
4 displays  $\lambda _{\max }(\mathrm{R}_{c})$ as a function
of the wave number $k$ and all  other parameters fixed. The eigenvalue
with maximum real part corresponds to $k=15$ \ and as it is purely real, the
bifurcation is stationary. Depending on the different parameters a great
variety of instabilities are obtained either stationary or oscillatory with
different growing  modes, which are analyzed next.

\subsubsection{Influence of $\protect\beta$ and $\protect\delta$}

In this section we study how the shape factor $\beta $ and the temperature
quotient $\delta $ both related to the bottom heat boundary condition affect
to the instabilities. To start with we fix parameters $\Gamma =10$, $\mathrm{%
Pr}=0.4$ and $B=0.05$ and take R as control parameter. 
Figures 5a and 5b display critical values of R$_{c}$ as a
function of $\delta $ for $\beta =0.5$ and $\beta =5$, respectively.
Critical wave numbers $k_{c}$ and the corresponding growing modes are shown
in these figures where void circles correspond to stationary bifurcations
while crossed ones stand for oscillatory instabilities. 
Both figures show regions with qualitatively different behaviors which are
separated by vertical lines. In figure 5a for $\delta >0.15$ and in
figure 5b for $\delta >0.055$ the critical R increases when $\delta 
$ decreases. At the $\delta $ limit values, the R threshold tends to grow so
much that we have not found it. This suggests that there is an asymptote
where the critical R$_{c}$ tends to infinity. As $\delta $ is increased, the
control parameter R decreases and so does the critical wave number.
Comparing figures 5a and 5b, which differ in its $\beta $
values, we notice more localized bifurcating structures for small values of $%
\beta $. This part of the diagram is called region I. At different $\beta $
the critical R$_{c}$ tends to infinity at different $\delta $ values. In
figure 6, the curve $C_{1}$ (solid line) that limits
region I marks the position of this asymptote in the $\beta -\delta $ plane.
An example of a basic state in region I is presented in Figure 3a).
It corresponds to a return flow.

In figure 5b, at $-0.05<\delta <0.055,$ an oscillatory bifurcation
is obtained with wave number $k=0$. This is detailed in figure 5c,
where a zoom of this region is displayed. The critical R$_{c}$ \ tends to
infinity as $\delta $ tends to the interval limits. The oscillatory
mode $k=1$ is just below the instability threshold but very close to it, at
the same parameter values. Figure 7 displays the values of $\lambda
_{\max }(\mathrm{R}_{c})$ as a function of the wave number $k$. The zero
mode is the preferred one, but as the mode one is so close to it,
it could be possible that both modes appear after the bifurcation. The mode $%
k=0$ corresponds to a target pattern (see figure 8a) and the mode $%
k=1$ to a giant one-armed spiral (see figure 8b). These plots
represent contours of the growing temperature field eigenfunction and their
structure are quite similar to those described in \cite%
{boden1,morris1,boden2,nature}. The zone where we find these behaviours is
called zone II (see figure 6). We do not find such instabilities
for $\beta =0.5$. In fact they exist only for large enough $\beta $ values
in a small interval around $\delta =0$, excluding $\delta =0$. This is
clearly depicted in Figure 6 where curve $C_{2}$ (dashed-
dotted line) limits region II. Figure 3b) displays an example of a
basic state in region II. It corresponds to a linear flow.

In figure 5a for $\delta <0.15$ and in figure 5b for $%
-0.186<\delta <-0.05,$ the bifurcations are stationary with a large wave
number which increases when $\delta $ decreases. The bifurcating structures
are localized at the outer part of the cylinder as figures 5a and 
5b display. In this range we can observe how for $\beta =0.5$ the
critical R$_{c}$ decreases with $\delta $ while for $\beta =5$\ R$_{c}$\
tends to grow to infinity at the limit $\delta =-0.186$. This suggests that
there is an asymptote where the critical R$_{c}$ tends to infinity. This
behavior corresponds to region III in the $\beta -\delta $ plane 
displayed in figure 6. Figure 3c) displays an example of
a basic state in region III.

Finally for $\delta <-0.186$ \ and $\beta =5$ \ no instability has been
found in zone IV. This means that in this case there exist external
parameter sets at which the axisymmetric basic state is very robust and does
not bifurcate to 3D structures. This regime is surrounded by curve $C_{3}$
(dashed line) in the $\beta -\delta $ diagram in figure 6. Figure 
3d) displays an example of a basic state in zone IV.

\subsubsection{Influence of the rest of the parameters}

\textit{Aspect ratio $\Gamma $.} Results in previous subsections are based
on a rather large aspect ratio domain ($\Gamma =10$). At smaller aspect
ratii the geometry of the domain has more influence on instabilities.
Typically for uniform heating at smaller aspect ratii instability thresholds
are larger than those obtained for larger aspect ratii and critical wave
numbers are smaller. Figure 9 shows how the instability thresholds
and wave numbers change with $\Gamma $. It is observed that the standard
rules just described for the case of uniform heating are mantained, although
now oscillatory instabilities are possible. The same type of instabilities
as those previously described are observed.
We notice that  in figure 9 for $\Gamma =3$ a one-armed spiral is the most unstable
mode while giant one-armed spirals described before, where found to be 
less  preferred than target patterns with $k=0.$

\noindent \textit{Pr number.} Similar behaviors to those we have just
described are also present for large and infinite Pr number. For instance
the giant one-armed oscillatory spiral appears as well in a similar range of
values of the parameters $\beta $ and $\delta $ (see figure 8c) and
with the same conditions, the mode $k=1$ oscillatory is just below the
instability threshold but very close to it, at the same parameter values%
\textsf{.}

\subsection{Discussion and conclusions}

We have studied how stationary and axisymmetric basic states bifurcate to
different 3D structures depending on the shape factor $\beta $ and
temperature ratio $\delta =\triangle T_{h}/\triangle T_{v}$ appearing on the
temperature boundary condition. The influence of horizontal and
vertical temperature differences on instabilities has been addressed in
numerous works \cite{smith,daviaud,burguete, mercier,pof,jpa,brief,pof2}; in our results a new heat parameter is introduced related to the
shape of the applied heat. We have found that at each $\beta $, different $%
\delta $ values define zones with qualitatively different instabilities.
Typically at small $\delta $ the basic state bifurcates to target and giant
spiral wave patterns. The appearance of these structures is quite similar to
those reported in experiments such as \cite{boden1,morris1,boden2,nature}
for the uniform heating case. The reason for this similitude might be the
presence of a mean flow in those experiments close to our basic state, that
could be caused by a non strictly uniform heat source. 
We find that even for
infinite Prandtl number spiral patterns are present. Larger negative $\delta 
$ values originates very localized states at the outer part of the domain.
Localized structures in the centre of the cell are found at small $\beta $
values and positive $\delta $ values. A decreasing aspect ratio increases
the instability threshold and decreases the critical wavenumber. This is in
agreement with the uniform case which corresponds to $\delta =0$.

\section*{Acknowledgements}

This work was partially supported by the Research Grants MCYT (Spanish
Government) BFM2003-02832, CCYT (JC de Castilla-La Mancha) PAC-05-005-01/02, 
{MEC (Spanish Government), MTM2004-00797},  SIMUMAT S-0505-ESP-0158 (Comunidad de Madrid) and by the University of
Castilla-La Mancha. AMM acknowledges MCYT (Spanish Government) for a Ram\'{o}n y Cajal Research Fellowship.

\newpage

\noindent \textbf{Figure captions}

\noindent Figure 1

\noindent Physical set-up.

\noindent Figure 2

\noindent Temperature boundary conditions at the bottom for $\Gamma
=10$; a) for $\delta =1$ and different values of $\beta $
; b) for $\beta =0.5$ and different values of $\delta $.

\noindent Figure 3

\noindent a) Isotherms and velocity field of the basic state corresponding
to values of the parameters $\mathrm{R}=356880$, $\Gamma =10$, $\mathrm{Pr}%
=0.4$, $B=0.05$, $\delta =0.25$ and $\beta =0.5$; b) isotherms and velocity
field of the basic state corresponding to values of the parameters $\mathrm{R%
}=19064$, $\Gamma =10$, $\mathrm{Pr}=0.4$, $B=0.05$, $\delta =0.05$ and $%
\beta =5$; c) isotherms and velocity field of the basic state corresponding
to values of the parameters $\mathrm{R}=30377$, $\Gamma =10$, $\mathrm{Pr}%
=0.4$, $B=0.05$, $\delta =0.15$ and $\beta =0.5$; d) isotherms and velocity
field of the basic state corresponding to values of the parameters $\mathrm{R%
}=10250$, $\Gamma =10$, $\mathrm{Pr}=0.4$, $B=0.05$, $\delta =-0.6$ and $%
\beta =5$.

\noindent Figure 4

\noindent Maximum real part of the growth rate $\lambda$ as a function of $k$ 
for basic state at threshold $\mathrm{R}_c=15095$, the rest of
the parameters are $\Gamma=10$, $\mathrm{Pr}=0.4$, $B=0.05$, $\delta=0.05$
and $\beta=0.5$. The maximum determines the critical $k_c=15$. Void
circles correspond to real eigenvalues while crossed ones stand for complex
eigenvalues.

\noindent Figure 5

\noindent a) Critical R, wave number $k$ values and growing modes as a
function of $\delta$ for $\beta=0.5$. Remaining external parameters are $%
\Gamma=10$, $\mathrm{Pr}=0.4$ and $B=0.05$; b) the same for $\beta=5$. Void
circles correspond to stationary instabilities while crossed ones stand for
oscillatory instabilities; c) A zoom of zones II and III of b). Dashed
vertical lines correspond to the asymptotes.

\noindent Figure 6

\noindent Different regions of bifurcations in the $\beta-\delta$ plane. The
rest of the parameters are $\Gamma=10$, $\mathrm{Pr}=0.4$ and $B=0.05$.

\noindent Figure 7

\noindent Maximum real part of the growth rate $\lambda$ as a function of $k$ 
for basic state at threshold $\mathrm{R}_c=19064$ corresponding
to the target ($k_c=0$) and the giant one-armed spiral ($k=1$) just below
the instability threshold but very close to it. The rest of the parameters
are $\Gamma=10$, $\mathrm{Pr}=0.04$, $B=0.05$, $\delta=0.05$ and $\beta=5$.
Void circles correspond to real eigenvalues while crossed ones stand for
complex eigenvalues.

\noindent Figure 8

\noindent a) Target at the instability threshold $\mathrm{R}=19064$. The rest of the parameters are 
$\Gamma =10$, $\mathrm{Pr}=0.4$, $B=0.05$, $%
\delta =0.05$, and $\beta =5$; b) giant one-armed spiral just below
the instability threshold but very close to it, at the same parameter
values; c) giant one-armed spiral for $Pr=\infty $, the rest of the
parameters are $\mathrm{R}=4674$, $\Gamma =10$, $B=0.2$, $\delta =0.1$ and $%
\beta =5$.

\noindent Figure 9

\noindent Critical R, wave number $k$ values and growing modes as a function
of $\Gamma $ for $\beta =0.5$, $\mathrm{Pr}=0.4$, $B=0.05$ and $\delta =1$.

\noindent \textbf{Table captions}

\noindent Table I

\noindent Critical Rayleigh numbers R$_{c}$ for different orders of
expansions in Chebyshev polynomials. The critical wave number is $k_{c}=0$
and the parameters are $\Gamma =10$, $Pr=0.4$, $\delta =0.05$, $B=0.05$ and $%
\beta =5$.

\noindent

\begin{center}
\begin{tabular}{|c|c|c|c|c|c|}
\hline
& $N=9$ & $N=11$ & $N=13$ & $N=15$ & $N=17$ \\ \hline
$L=19$ & $19069$ & $19070$ & $19070$ & $19070$ & $19070$ \\ 
$L=25$ & $19055$ & $19056$ & $19056$ & $19056$ & $19056$ \\ 
$L=29$ & $19065$ & $19066$ & $19066$ & $19066$ & $19066$ \\ 
$L=33$ & $19063$ & $19064$ & $19064$ & $19064$ & $19064$ \\ 
$L=37$ & $19063$ & $19064$ & $19064$ & $19064$ & $19064$ \\ \hline
\end{tabular}
\end{center}

\newpage
\begin{center}
{\it Figure 1}\\
\begin{tabular}{c}
\epsfig{file=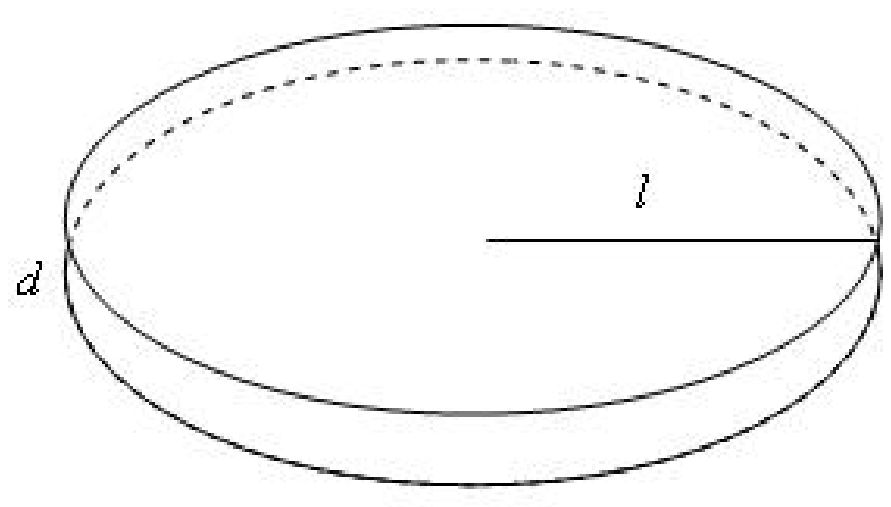,width=12cm} \\
\textit{}%
\end{tabular}
\end{center}
\newpage
\begin{center}
{\it Figure 2}\\
\epsfig{file=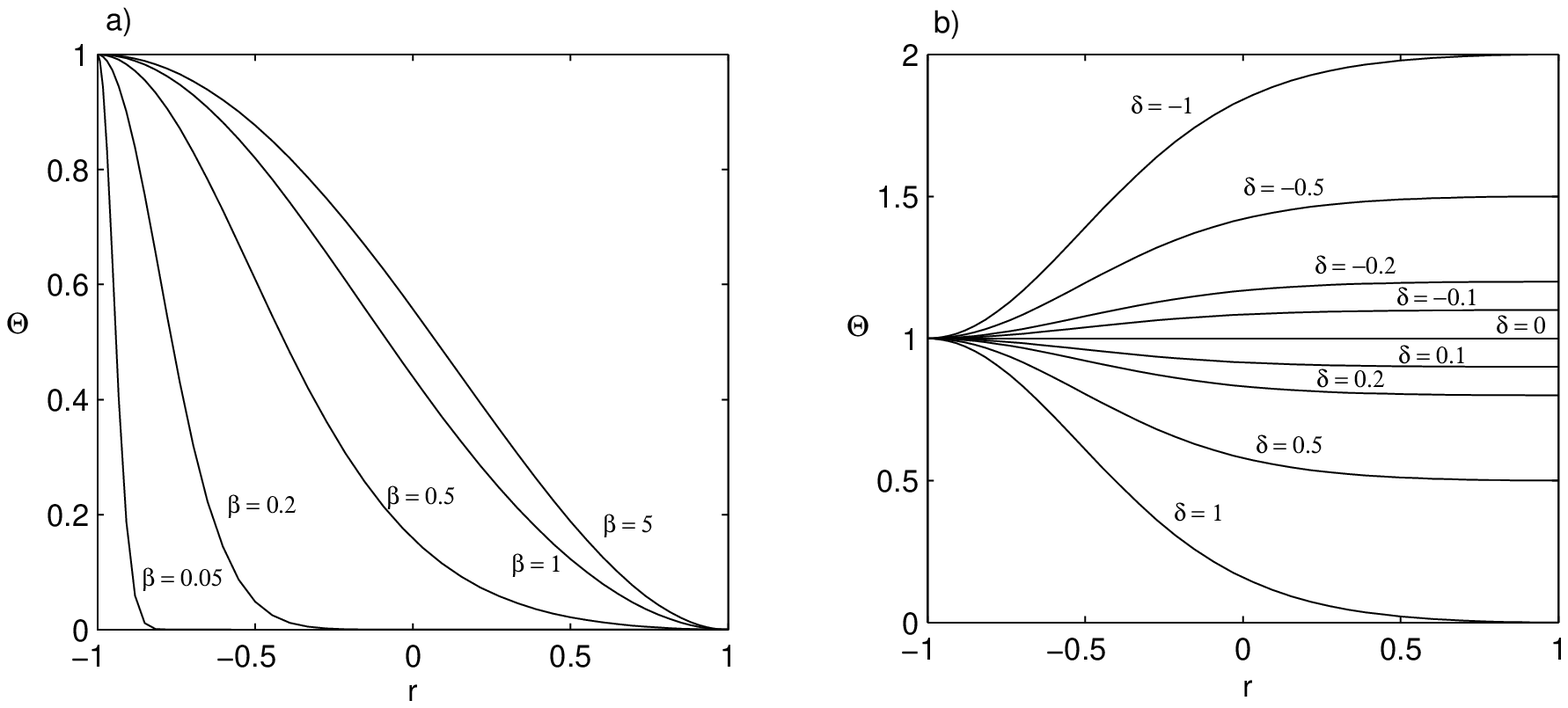,width=16cm} \\
\textit{}%
\end{center}
\newpage
\begin{center}
{\it Figure 3}\\
\epsfig{file=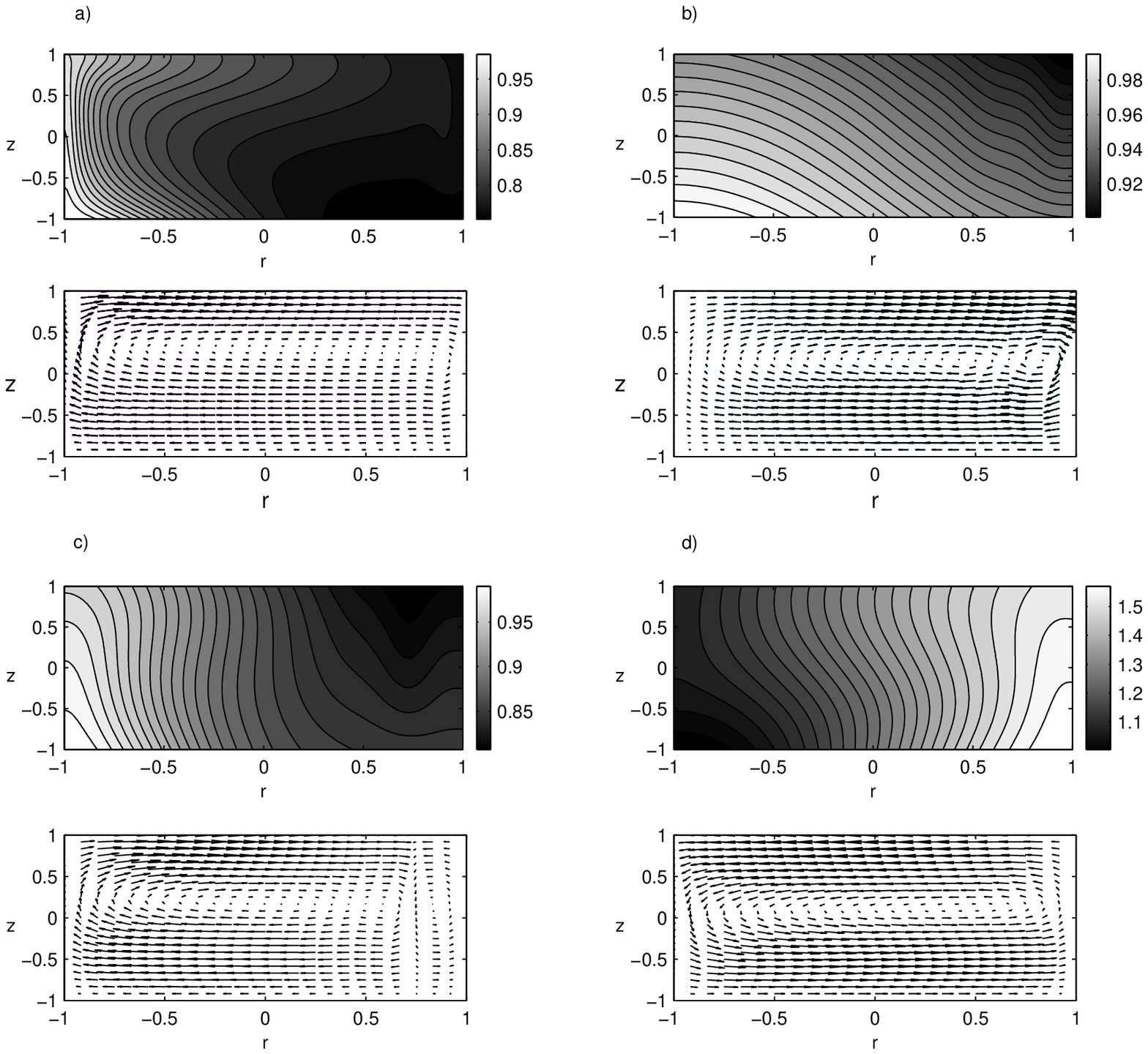,width=16cm} \\
\textit{}%
\end{center}
\newpage
\begin{center}
{\it Figure 4}\\
\epsfig{file=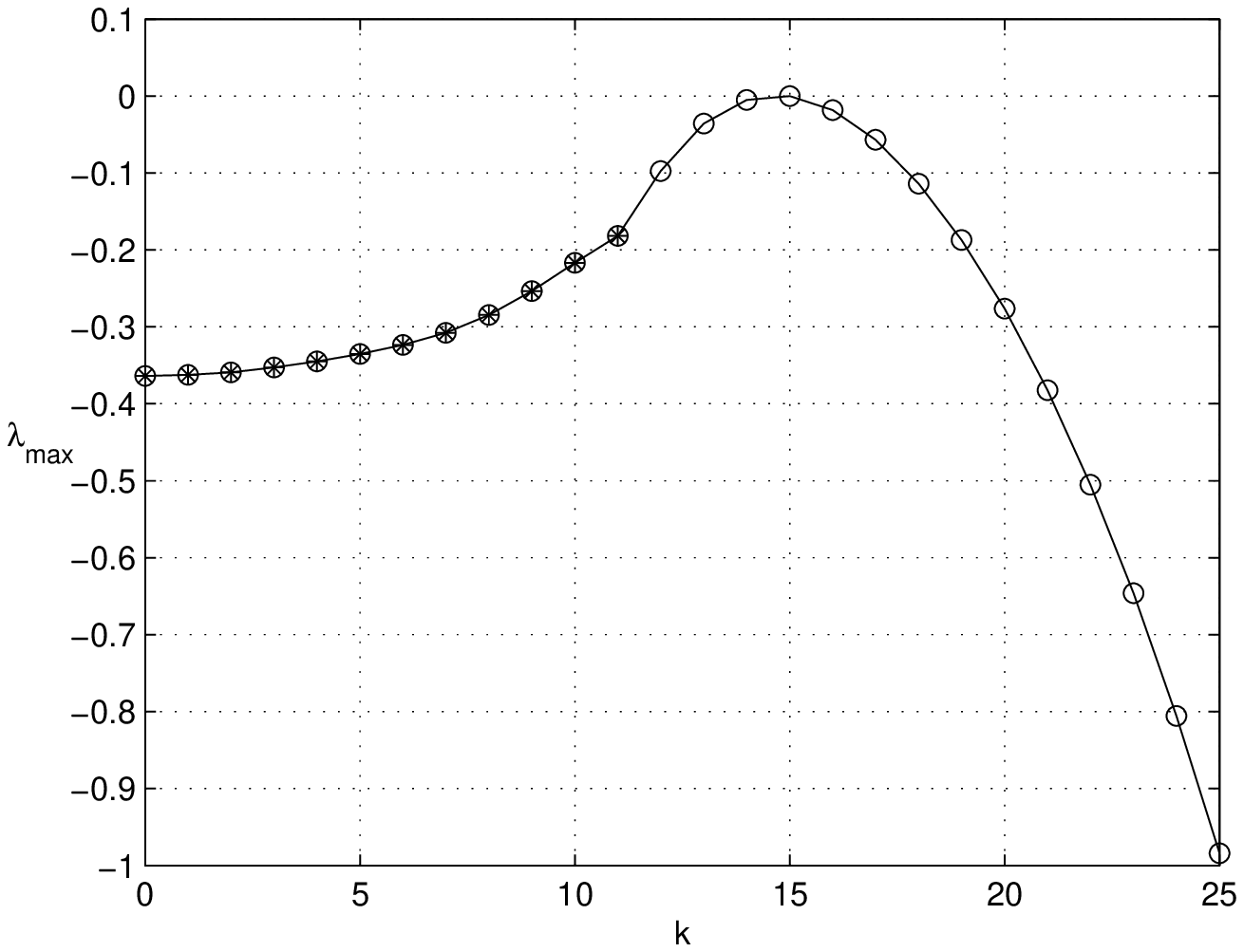,width=16cm} \\
\textit{}%
\end{center}
\newpage
{\it Figure 5}\\
\epsfig{file=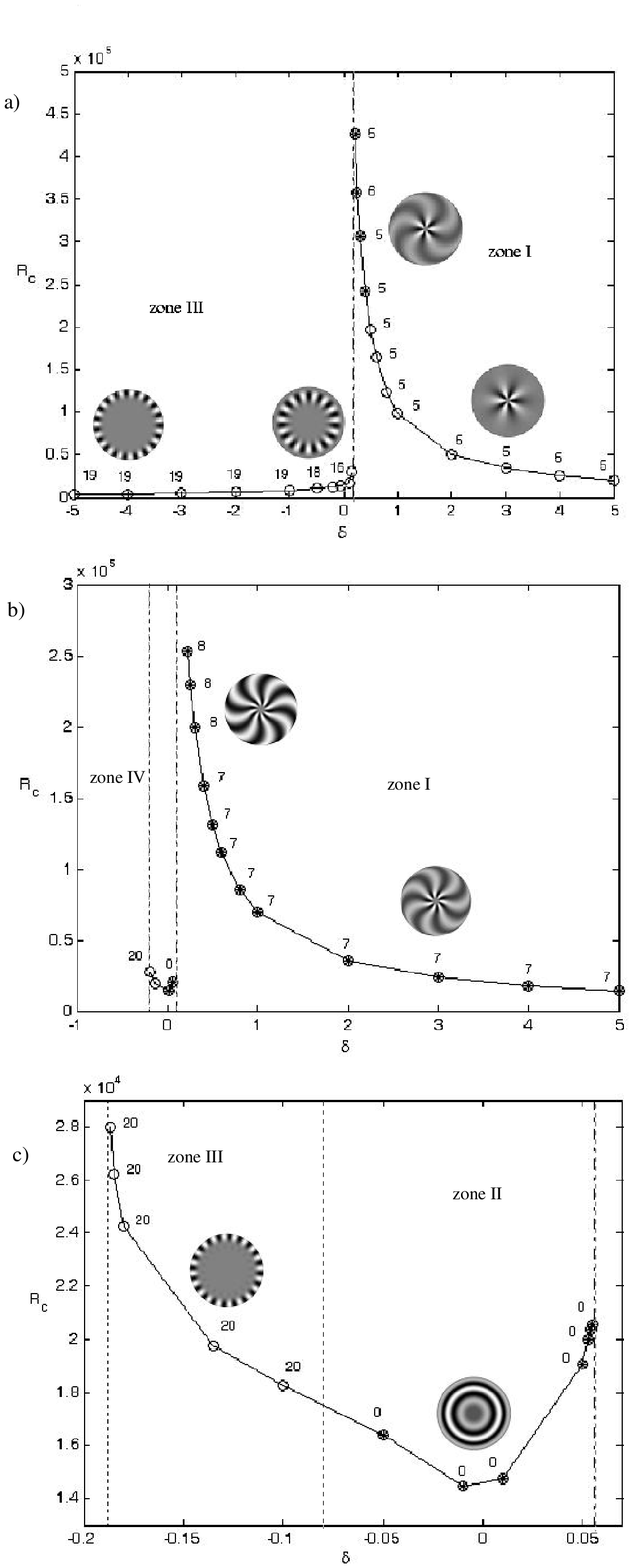,width=8cm} \\
\textit{}%
\newpage
\begin{center}
{\it Figure 6}\\
\epsfig{file=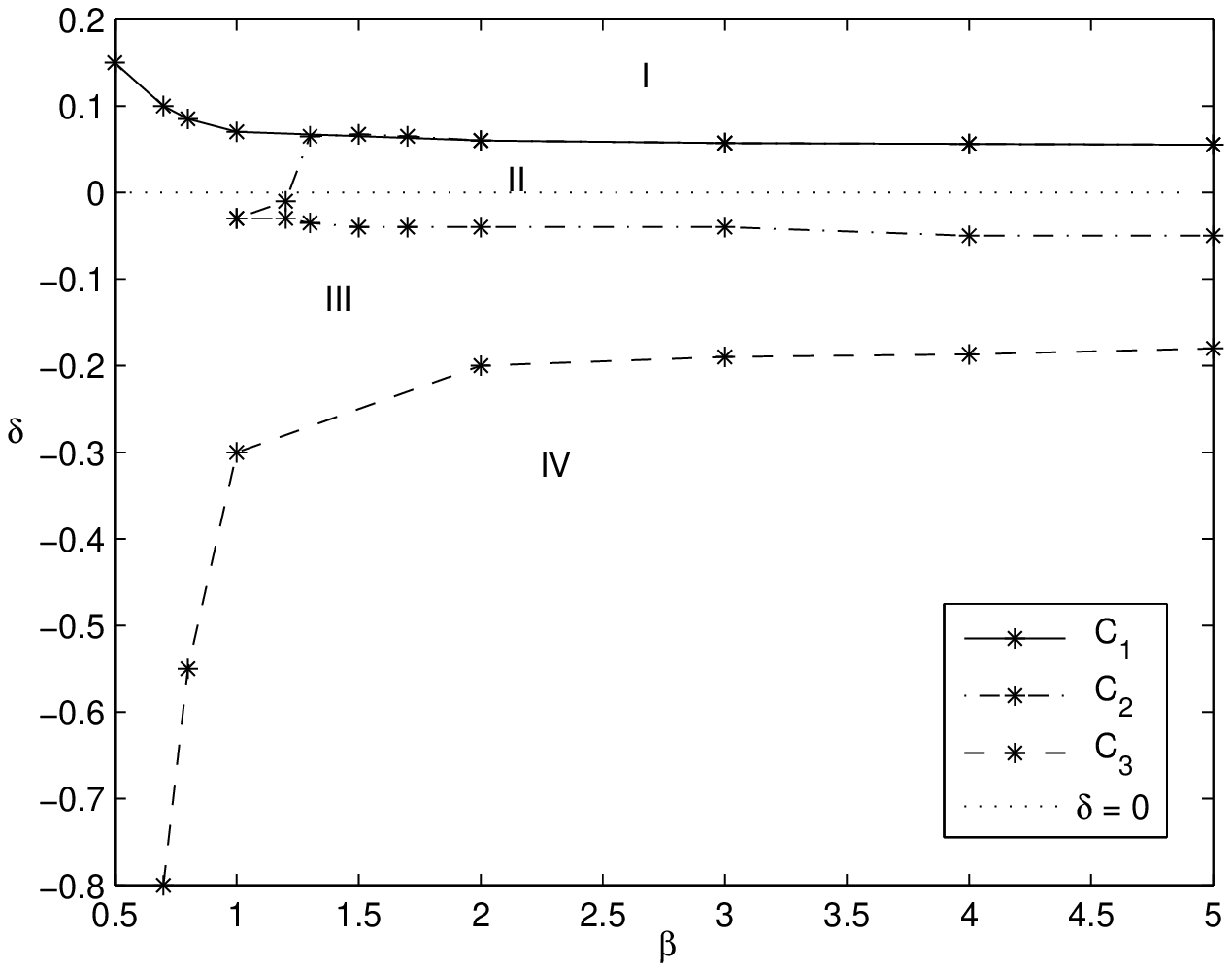,width=16cm} \\
\textit{}%
\end{center}
\newpage
\begin{center}
{\it Figure 7}\\
\epsfig{file=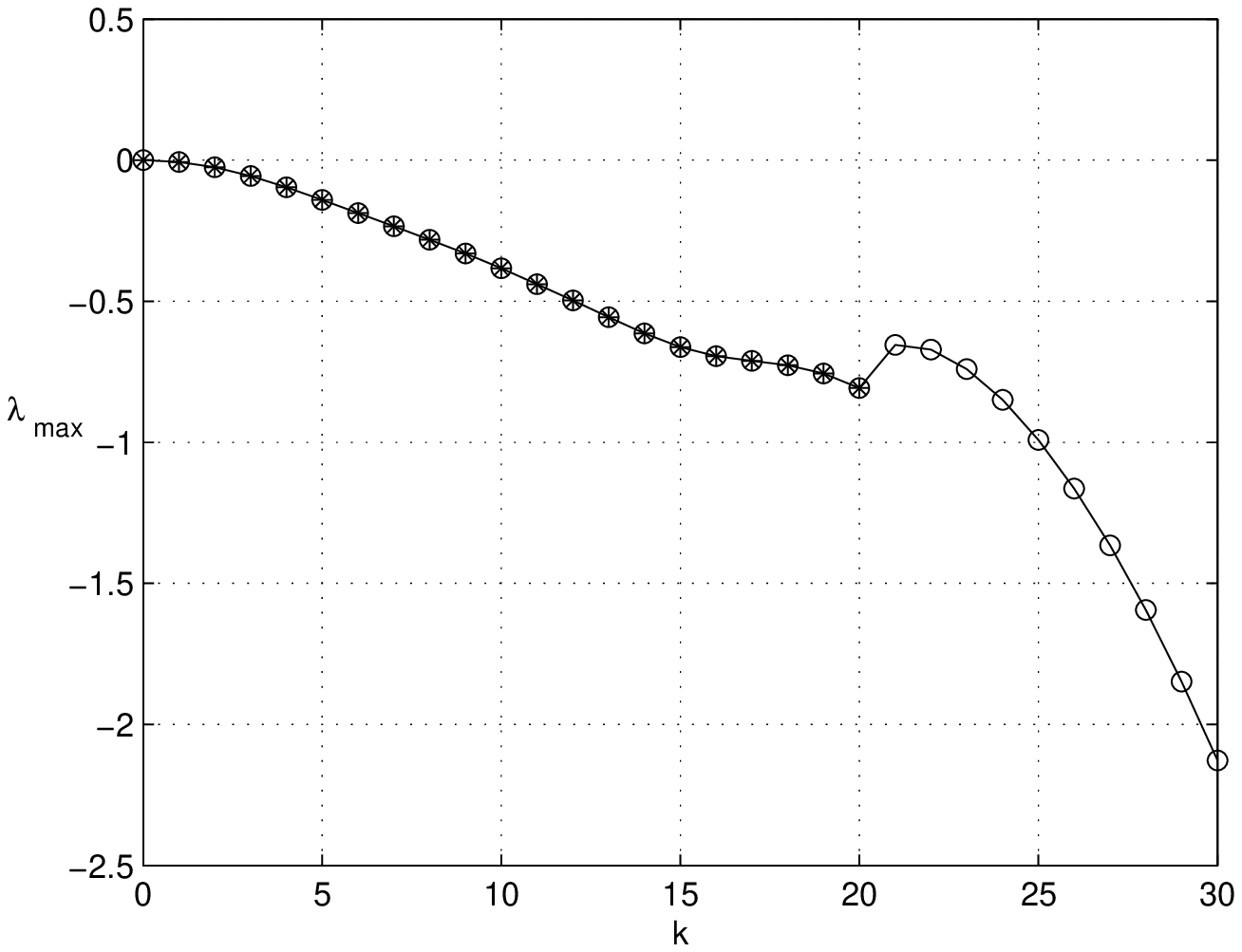,width=16cm} \\
\textit{}%
\end{center}
\newpage
\begin{center}
{\it Figure 8}\\
\epsfig{file=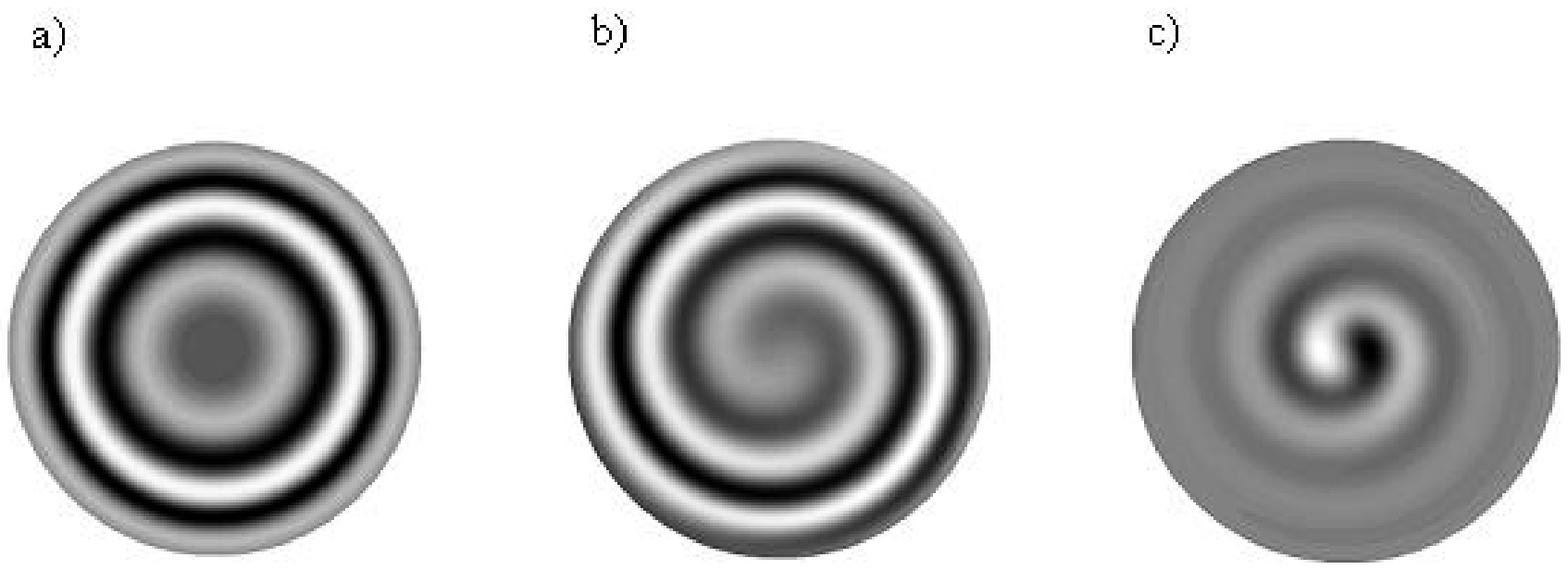,width=16cm} \\
\textit{}%
\end{center}
\newpage
\begin{center}
{\it Figure 9}\\
\epsfig{file=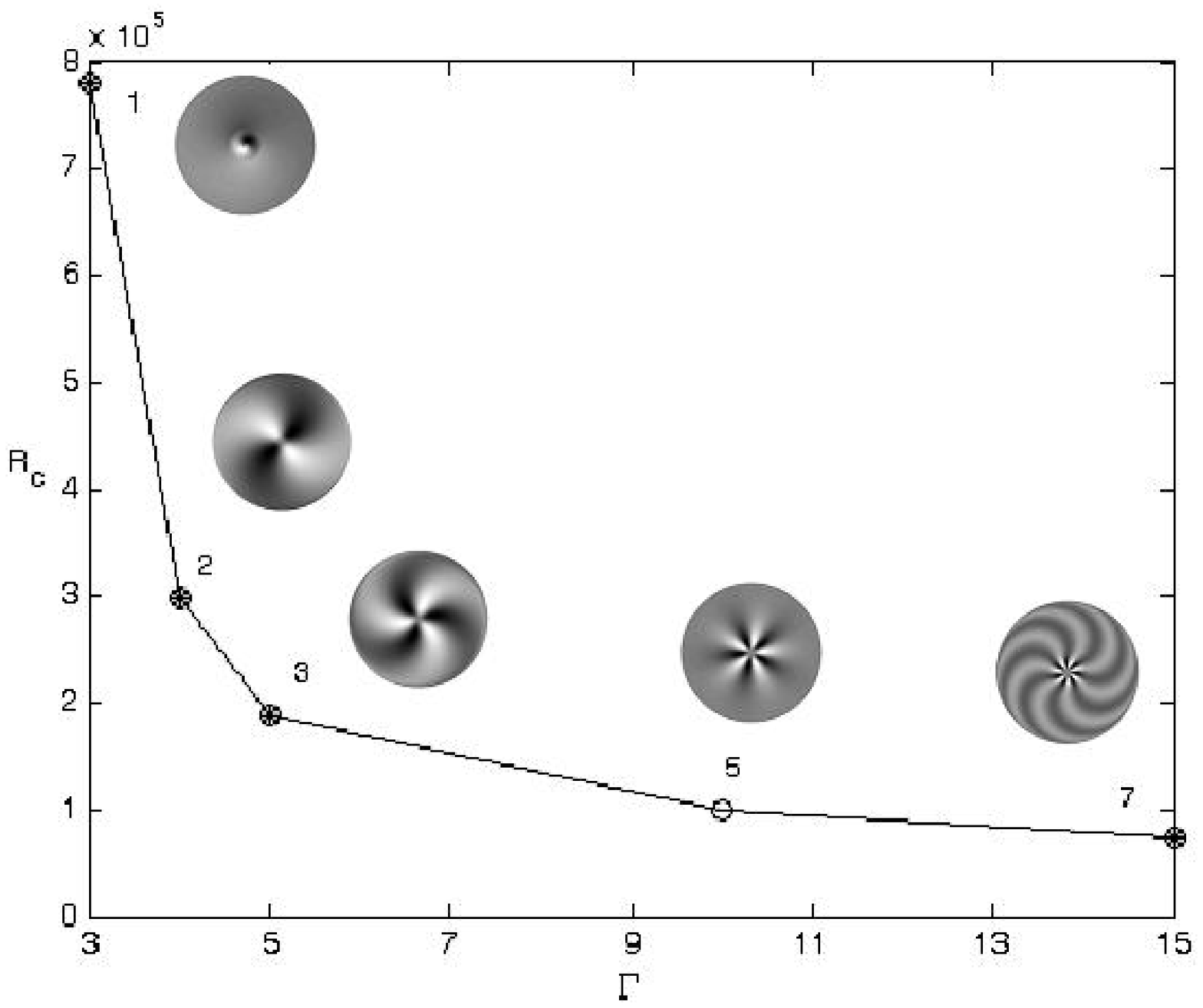,width=16cm} \\
\textit{}%
\end{center}

\end{document}